# A study of the fractional quantum Hall effect: Odd and even denominator plateaux

M. A. Hidalgo

Departamento de Física y Matemáticas

Universidad de Alcalá

Alcalá de Henares, Madrid, Spain

Correspondence and request for materials should be addressed to miguel.hidalgo@uah.es

## Abstract

We present a different approach to the fractional quantum Hall effect (FQHE), focusing it as a consequence of the change in the symmetry of the Hamiltonian of every electron in a two-dimensional electron gas (2DEG) under the application of a magnetic field and in the presence of an electrostatic potential due to the ionized impurities, and leading to a breaking of the degeneration of the Landau levels. As the magnetic field increases the effect of that electrostatic potential evolves, changing in turn the spatial symmetry of the Hamiltonian: from continuous to discrete one. The aim of both works is to give a different picture not only of the FQHE phenomenon, but a coherent one with the integer quantum Hall effect (IQHE) and consistent with the model already described in Hidalgo[7, 8, 9]. Therefore the model gives a global view of both effects, showing that they are aspects of the same phenomenon, and justifying not only the appearance of the odd denominator plateaux but also the even ones; and giving some physical reasons for the experimental fact that there are much more odd than even denominator plateaux, hardly observed.

## Introduction

One of the amazing phenomena in last decades in solid state physics have been the appearance of the integer quantum Hall effect (IQHE), in 1980[1, 2], and the FQHE two years later, in 1982[3]. Both phenomena, phenomenologically similar, show their main features in the data related to magnetotransport in a two-dimensional electron system (2DES): minima or zeroes in the longitudinal resistance, i.e. Shubnikov-de Haas effect (SdH), and well-defined plateaux in the Hall resistance (at integer or fraction values of the fundamental Hall resistance $R_H = h/e^2$). Although the IQHE can be observed in a 2DESs form in quantum wells (QW), MOSFET or semiconductor heterostructures, to measure the FQHE high motilities heterostructures (like GaAs-AlGaAs) are necessary to the corresponding plateaux in the Hall measurements be clearly observed.

Theoretically, while the IQHE may be understood from Landau quantization as a single particle localization effect[4], however, the FQHE is thought to be a consequence of the strong electronic correlations when a Landau level is only partially filled and the Coulomb interaction between the electrons of the gas became relevant. In 1983, Laughlin[5] proposed that the origin of the observed FQHE 1/3, -as well any $1/q$ with $q$ being an odd integer-, is due to the formation of a correlated incompressible electron liquid; and in that view the electron-electron interaction is analyzed constructing several electron wavefunctions *ad hoc*, i.e. therefore the main task is searching for explicit trial wave functions corresponding to the states of the 2DES that does not break any continuous spatial symmetry and show energy gaps. The nature of these states is associated with uniform density condensates. After the discovery of the 1/3 plateau, many other fractional plateaux have been observed, globally grouped in the general expression $p/(2sp\pm1)$, with $s$ and $p$ being integers. This series has been interpreted in the context of the so-called composite fermions, according to which the FQHE may be viewed as an IQHE of quasiparticles consisting of an electron capturing an even number of flux quanta[6]. These composite entities become elementary particles of the system.

The observation of the FQHE was completely unexpected, initially with the appearance of the 1/3 plateau[2], and then confirmed with the observation of large series of other fractional

plateaux, most of them odd: {4/5, 2/3, 3/5, 4/7, 5/9, 5/3, 8/5, 11/7, 10/7, 7/5, 4/3, 9/7}[10]; {5/3, 8/5, 10/7, 7/5, 4/3, 9/7, 4/5, 3/4, 5/7, 2/3, 3/5, 4/7, 5/9, 6/11, 7/13, 6/13, 5/11, 4/9, 3/7, 2/5, and also 8/3, 19/7, 33/13, 32/13, 7/3, 16/7}[11,12]; {2/3, 7/11, 3/5, 4/7, 5/9, 6/11, 5/11, 4/9, 3/7, 2/5, 1/3, 2/7, 3/11, 4/15, 3/13, 2/9}[13]; {14/5, 19/7, 8/3, 13/5, 23/9, 22/9, 17/7, 12/5, 7/3, 16/7, 11/5}[14]; {19/5, 16/5, 14/5, 8/5, 7/3, 11/5, 11/3, 18/5, 17/5, 10/3, 13/5, 12/5}[15]. Although sometimes showing the measurements even planteaux: {15/4, 7/2, 13/4, 11/4, 5/2, 9/4}[10]; {11/4, 5/2, 9/4}[11]; {1/4}[13]; {11/4, 21/8, 5/2, 19/8, 9/4}[14]; {7/2, 5/2}[15].

Looking at the extensive series observed, and their recurrence in the experiments on different 2DESs, the fundamental character of the FQHE phenomenon is obvious, seeming clear the common origin of the physics underneath. In this context, it is important to highlight the fact that in all FQHE measurements the integer plateaux are also observed.

From a detailed analysis of the experimental data related to the FQHE published in literature, we can establish the following set of general features found in all of them: 1) Basically the main fractions observed in the measurements correspond to odd plateaux; 2) the electron densities are always low; 3) the 2DESs have high mobilities; 4) the integer plateaux are always observed and, even more, much more well defined than in the 2DESs where only the IQHE appears.

The two last experimental facts, 3) and 4), seems to contradict the Laughlin argument for the IQHE[2], concerning on the need of the presence of impurities to explain the appearance of the integer plateaux –in the FQHE measurements, all of them on high mobility samples, the integer plateaux are extremely well defined-. Moreover, the observation of both, integer and fractional plateaux, in any FQHE experiment stand in contradiction with the theoretical view give in literature for each one: the IQHE as a phenomenon associated with a 2DEG, while the FQHE related to a Fermi liquid (Laughlin's picture). In sight of the experimental results, these pictures would imply a succession of phase transitions in the 2DES, (electron gas-liquid transitions).

Indeed, several attempts to understand the SdH oscillations and the IQHE have been published[2, 7, 8, 9]. The most accepted picture is based on the 'gendanken' experiment thought up by Laughlin[2], where the 2DES localized states due to impurities and defects would play a crucial role to explain the plateaux of the IQHE and the simultaneous minima values, close to zero, in the SdH. However, as we have mentioned above, the main criticism to this idea is the experimental evidence: higher electron mobility provides better plateaux precisions. But, alternatively, the model for both, the IQHE and the SdH, described in reference 7, (and also in 8 and 9), does not precise involving localized states to justify the presence of plateaux in the IQHE, and zeroes in the SdH effect, but just two assumptions: a constant Fermi level -fixed by the 3D environment where the 2DES-is immersed, and, secondly, the possibility of existence of a flow of electrons from/to the QW to/from the doping zone of the heterostructure -large relative variations in the 2DES electron density are possible with negligible variations in the 3D doping area electron density-.

Therefore, the view we propose implies that both phenomena, IQHE and FQHE, are aspects of the same phenomenon that, based on the quantization of the states of the 2DEG, will only depend on the external fields undergone by the electrons of the gas. Thus, the starting point for our FQHE study will be an extension of the model we develop for the IQHE[7, 8, 9], and then analyzing the FQHE in the context of a single particle approach.

**Study of the experimental maxima of the SdH oscillations**

The Fermi level of a 2DEG is given by the equation

$$E_F = \frac{2\pi\hbar^2 n_e}{m^*} \quad (1)$$

where $n_e$ is the electron density of the 2DEG (easily obtained from the Hall measurements at low magnetic fields), and $m^*$ the corresponding effective mass.

On the other hand, from the experimental SdH oscillations of any FQHE observation, one can check that the maxima of every oscillation match with the values of the magnetic field given by the relation

$$B_{q,p} \approx \frac{q}{p}\frac{m^* E_F}{\hbar e} \tag{2}$$

with $p$ and $q$ natural numbers[7], and $E_F$ given by (1). Even more, each fraction $p/q$ corresponds to the plateaux with the same fractional number. In fact, the same expression (2) is valid to determine the position of the maxima of the SdH oscillations in the experiments of the IQHE. Hence, because the maxima of the SdH directly reflects the energy states of every electron of the 2DEG, then, we conclude that, in the same way as the SdH maxima in any IQHE experiment are consequence of the existence of the quantization in Landau levels -determined by integer values-, we can assume that the SdH maxima in any FQHE experiment are due to the quantization in fractional levels (determined by $p/q$).

**Origin of fractional odd denominator quantum levels**

Therefore, following the idea just given in the last paragraph of previous section, we try to find the cause of such quantization in fractional states of energy, of course, basing us on the single electron approximation. Our starting point will be the symmetric gauge $\vec{A} = B(-y,x)/2$, being $\vec{B} = (0,0,B)$ the applied external magnetic field, assumed perpendicular to the plane defined by the 2DES. The Hamiltonian in the effective mass approximation is given by the expression

$$H_1 = \frac{1}{2m^*}\left(\vec{p} + e\vec{A}\right)^2 \tag{3}$$

And the wave functions obtained for this Hamiltonian from the Schrödinger equation is

$$\psi_n^m = \left[2^m \frac{(m+n)!}{n!}\right]^{-\frac{1}{2}} (x'-iy')^m \, exp\left(-\frac{r'^2}{4}\right) L_n^m\left(\frac{r'^2}{2}\right) \tag{4}$$

being $x' = x/R$, $y' = y/R$, $r'^2 = x'^2 + y'^2$, $R = \sqrt{\hbar/eB}$ the magnetic length and $L_n^m\left(\frac{r'^2}{2}\right)$ the Laguerre polynomials. This set of wave functions is orthonormal each other respect to both, the $n$ and $m$ indexes, this last one associated with the angular momentum of the electron.

The energy states correspond to the Landau levels, i.e.

$$E_n = \left(n + \frac{1}{2}\right)\hbar\omega_0 = (2n+1)E_0 \tag{5}$$

$n$=0, 1, 2…, $\omega_0 = eB/m^*$ the fundamental angular frequency, and $E_0 = \hbar\omega_0/2$. These levels are degenerate in all possible angular momentum states, determined by $m$.

As it is well-known, the expected value of the square of the distance from the center of the trajectory of the electron to the origin -in our case each ionized impurities-, which we will reference below as Larmor radius, $R_O$, is given by

$$< R_O^2 >= qR^2 \tag{6}$$

being $q$ an odd number. This equation (6) allows us consistently defining the wavelength of the electron through the relation $\lambda_q = 2\pi\sqrt{< R_O^2 >}$ . Then, its corresponding wave number will be

$$k_q = \frac{2\pi}{\lambda_q} = \frac{1}{\sqrt{< R_O^2 >}} = \frac{1}{\sqrt{q}}\frac{1}{R} = \frac{1}{\sqrt{q}}\sqrt{\frac{eB}{\hbar}} \tag{7}$$

But, actually, the general Hamiltonian of every electron in the 2DEG is

$$H_2 = \frac{1}{2m^*}\left(\vec{p} + e\vec{A}\right)^2 + U(\vec{r}) = \frac{1}{2m^*}\left(\vec{p} + e\vec{A}\right)^2 + U^i(\vec{r}) + U^e(\vec{r}) \tag{8}$$

The last two terms correspond to the energy contribution of the electrostatic potentials due to the ionized impurities, $U^i(\vec{r})$, and the electron-electron interaction, $U^e(\vec{r})$. (Later we discuss about the Zeeman term.) Of course, in all below we assume that the electrostatic potentials term is a perturbation respect to the predominant effect of the magnetic field.

At low magnetic field the Larmor radius of every electron is large; then their interactions $U^i(\vec{r})$ with the ionized impurities, -assuming they are distributed with a mean distance $d_i$ (value determined by the density of impurities), and because in that case we have $d_i << 2R_O$, $R_O$ given by Equation (6)-, can be assumed to be a uniform term, i.e. $U_0^i \approx < U^i(\vec{r}) >$ along the 2DES. On the other hand, the electrostatic potential associated with the electron-electron interaction can be neglected in average due to the symmetry related to the electron distribution in the 2DES. Therefore, we suppose that the global electrostatic potential term acting on every electron is given by $U_0 \approx U_0^i$. Thus, under these conditions, the energy states for every electron are $E_n = (2n+1)E_0 + U_0$.

However, at high magnetic fields the Larmor radius will be of the order of $d_i$, and, then, the effect of the ionized impurities over every electron will now contribute to the Hamiltonian with a non-uniform term $U^i(\vec{r})$. But this evolution in the effect of the electrostatic interaction term with the increasing magnetic field involves a change in the symmetry of the Hamiltonian of the 2DEG, changing from the initial continuous spatial symmetry, determined by $U_0$, to a discrete one, $U^i(\vec{r})$, determined by the distribution of the ionized impurities. Therefore, under this new condition, we can view the new states associating them with an arrangement of cyclotron orbits reflecting the symmetry in the distribution of the ionized impurities: a short-range order like shown in Figure 1, where with black dot points are represented the ionized impurities. To characterize this new symmetry we establish some correlation lengths related to that short-range order, which we will express as $\eta d_i$. This term $\eta d_i$ gives us the most probable spatial distance between the centers of the electron orbits in the 2DEG between closest neighbor electrons. In fact, as a consequence of that, it is hoped for the breaking of the degeneration of each Landau level.

Then, the new symmetry requires that $H_2(r) = H_2(r + \alpha\eta d_i)$, where $\alpha$ is a natural number. And we express the correlation through the relation

$$<\psi_n^m(r + \alpha\eta d_i)/H_2/\psi_n^m(r + [\alpha \pm 1]\eta d_i)> = \pm\gamma/2 \qquad (9)$$

where we suppose that $\gamma$ is the same for all cyclotron orbits. (We also assume the higher correlations terms negligible.) Although the set of functions $\{\psi_n^m\}$, (4), does not verify the Bloch theorem, we can construct a new base reflecting that new short range order, taking the linear combination of the cyclotron orbits functions, i.e.

$$\Phi_{d_i}(k_q, r) = \frac{1}{\sqrt{N}} \sum_{\alpha=1}^{N} exp(ik_q \cdot \alpha\eta d_i)\psi_n^m(r - \alpha\eta d_i) \qquad (10)$$

being $N$ the normalization term representing the number of cyclotron orbits in any direction of the 2DES; this equation represents a set of orthonormal functions. Then, we have $\Phi_{d_i}(k_q, r + \beta\eta d_i) = exp(ik_q \cdot \beta\eta d_i)\Phi_{d_i}(k_q, r)$, and the dispersion relation $E = <\Phi_{d_i}(k_q, r)/H_2/\Phi_{d_i}(k_q, r)>$. Therefore, the new energy states are expressed by the equation $E = E_n \mp \gamma\, cos(k_q\eta d_i)$, that, can be approached, under the conditions assumed, as $E = E_n \mp \gamma \pm \frac{\gamma}{2}(k_q\eta d_i)^2$, from where it is easy to deduce that $\gamma = U_0$. Doing $\hat{E}_n = E_n \mp U_0$, we can write $E = \hat{E}_n \pm \frac{\gamma}{2}(k_q\eta d_i)^2 = \hat{E}_n + \hat{E}_q^\eta$, with

$$\hat{E}_q^\eta = \pm\eta^2 \frac{U_0}{2}(k_q d_i)^2 \qquad (11)$$

This equation (11) is similar to that corresponding to the free electron system with an effective mass $m^* = \frac{\hbar^2}{|U_0| d_i^2}$. Hence, equation (11), using equation (7), can be written as

$$\hat{E}_q^\eta = \pm \frac{\eta^2}{q} \frac{\hbar\omega_0}{2} = \pm \frac{\eta^2}{q} E_0 \qquad (12)$$

with $q$ being an odd number. Hence, the possible energy states for every electron of the 2DEG are

$$E = \hat{E}_n \pm \frac{\eta^2}{q} E_0 = (2n+1)E_0 \pm \frac{\eta^2}{q} E_0 + U_0 = \left(2n+1 \pm \frac{\eta^2}{q}\right) E_0 + U_0 \qquad (13)$$

The expected correlations in the arrange of the electron cyclotron orbits in a 2DES is like that as shown in Figure 1, i.e. short range order to the first and second neighbors, as explained above, and which correspond to values of the correlation length, $\eta d_i$, with $\eta$=1 and $\sqrt{3}$, respectively. In fact, we can in general write $\eta = \sqrt{p}$, being the correlation index $p$ an odd number, (in the case of Figure 1, correlation indexes $p$=1 and 3 are drawn). Hence, we have the energy states for every electron

$$\frac{E}{\hbar\omega_0} = \frac{1}{2}\left(2n+1 \pm \frac{p}{q}\right) \qquad (14)$$

In Table I we detail these energy states for Landau levels $n$=0, 1 and 2, $q$ values between 3 and 13, and $p$=1 and 3. The fractions already observed are highlighted in red color. As it is seen, all those values coincide with the families of odd denominator, and their corresponding sequences, obtained in the experiments of the FQHE[10, 11, 12, 13, 14, 15].

Eventually, we have to take into account the contribution of the Zeeman and the spin-orbit coupling terms. These can be summarized in the expression $E_{spin} = g^* m^* \hbar\omega_0 / 4m$, being $g^*$ the generalized gyromagnetic factor. In the FQHE conditions we can consider that all the electrons are uniformly polarized.

In order to illustrate the model just presented we simulate both magnetoconductivities (and, then, the SdH and Hall effects) for a 2DEG. A detail description of the procedure for the IQHE is written in references 7, 8 and 9, and now we only summarize it. Firstly we have to obtain the density of states for an odd denominator, $q$, and for that purpose we use the Poisson sum formula, obtaining[7]

$$g_q(E) = g_0\left\{1 + 2\sum_{p=1}^{\infty} A_{\Gamma,p} A_{S,p} \cos\left[\left(\frac{2\pi p q E}{\hbar\omega_0} - \frac{g^*}{4}\right)\right]\right\} \quad (15)$$

where $A_{S,p} = \cos\left(\pi p \frac{g^*}{2}\frac{m^*}{m}\right)$ is the term associated with the spin and spin-orbit coupling,

and $A_{\Gamma,p} = \exp\left\{-\frac{2\pi^2 p^2 \Gamma^2}{\hbar^2\omega_c^2}\right\}$, the gaussian term related to the width of the energy levels, and due to the interaction of electrons with defects and impurities[16]. (For sake of simplicity we have assumed gaussian width for each energy level, and independent of the magnetic field.) From the density of sates the electron density is easily obtained[7, 8],

$$n = n_0 + \delta n = n_0 + \frac{2eB}{hq}\sum_{p=1}^{\infty}\frac{1}{\pi p} A_{\Gamma,p} A_{S,p} sen\left[X_F - \frac{g^*}{4}\right] \quad (16)$$

with $X_F = (2\pi p\, qE_F\ \hbar\omega_0 - g^*\ 4)$. (In this expression we have supposed a very low temperature, the ideal experimental condition to observe the FQHE.) From equations (15) and (16) the magnetoconductivities can be calculated[7, 8, 9].

For testing and comparing our model we have used the already classical experimental measurements by Willet et al.[11]. In Figures 2(a) and (b) we show the results at high magnetic fields for the Hall magnetoconductivity and both magnetoresistivities corresponding to the series with odd denominators $q$=3 and 5, respectively.

## Origin of fractional even denominator quantum levels

We now try to find the origin of the quantization in fractional states of energy with even denominator. In the previous section, from the expected value of the square of the distance from the center of the electron trajectory to the own origin for $q \geq 3$, $q$ being always an odd number, we have explained the appearance of the odd denominator plateaux. However, the analysis of the odd value $q=1$ remained pending. In this case the Larmor radius $R_O$ is given through the relation $< R_O^2 >= R^2$; this is its minimum possible value because of the fact that the coordinates determining the distance from the center of the electron trajectory to the own origin do not commute (on the other hand, this also coincides with the value of the cyclotron radius of every electrón). Hence, following the same arguments used in the previous section, this relation allows us defining consistently the wavelength of every electron in these states through the equation $\lambda_1 = 2\pi\sqrt{< R_O^2 >} = 2\pi R$ , and, then, its corresponding wave number

$$k_1 = \frac{2\pi}{\lambda_1} = \frac{1}{R} = \sqrt{\frac{eB}{\hbar}} \qquad (17)$$

In this case the change in the symmetry of the Hamiltonian of the 2DEG is also determined by the distribution of the ionized impurities. But to characterize this new symmetry under the conditions fixed, we again establish some correlation lengths related to that discrete distribution of the cyclotron orbits, which we will express as $\eta d$, $d$ being a fundamental distance. (This term gives us the most probable spatial distance between the centers of the electron orbits in the 2DEG between closest neighbor electrons.) As a consequence of that, it is hoped for the breaking of the degeneration of each Landau level.

Therefore, to impose the new symmetry implies $H_2(r) = H_2(r + \alpha\eta d)$, with $\alpha$ a natural number and, thus, the correlation can be required through the relation between nearest neighbors given by

$$< \psi_n^m (r + \alpha\eta d) / H_2 / \psi_n^m (r + [\alpha \pm 1]\eta d) >= \pm\gamma/2 \qquad (18)$$

where we suppose that $\gamma$ is the same for all cyclotron orbits. (We also assume the higher correlations terms negligible.) We can construct a new base reflecting that new short range order, taking the linear combination of the cyclotron orbits functions, i.e.

$$\Phi_d\left(k_1, r\right) = \frac{1}{\sqrt{N}} \sum_{\alpha=1}^{N} exp\left(ik_1 \cdot \alpha\eta d\right)\psi_n^m\left(r - \alpha\eta d\right) \tag{19}$$

similar to equation (10). Then, following the steps given in the previous section, we find that the new energy states can be approached under the conditions assumed as $E = \hat{E}_n \pm \frac{\gamma}{2}\left(k_1\eta d\right)^2 = \hat{E}_n + \hat{E}_1^\eta$, with

$$\hat{E}_1^\eta = \pm\eta^2 \frac{U_0}{2}\left(k_1 d\right)^2 \tag{20}$$

where we have again assumed an effective mass $m^* = \frac{\hbar^2}{|U_0| d^2}$. Hence, equation (20), using equation (17), can be written as

$$\hat{E}_1^\eta = \pm\eta^2 \frac{\hbar\omega_0}{2} = \pm\eta^2 E_0 \tag{21}$$

And the possible energy states for every electron of the 2DEG are

$$E = \hat{E}_n \pm \eta^2 E_0 = \left(2n+1\right)E_0 \pm \eta^2 E_0 + U_0 = \left(2n+1 \pm \eta^2\right)E_0 + U_0 \tag{22}$$

The expected correlations in the arrange of the electron cyclotron orbits in a 2DES under the conditions assumed in this section can be like that as shown in Figure 3, i.e. short range order to the first and second neighbors, where with black dot points are represented the ionized impurities, and in blue the cyclotron orbits, (the drawn square is only a guide for the eyes). As it is seen in the own figure, the correlation length $\eta d$ are such that $\eta = 1/\sqrt{2}$ and $\eta = 1/2$. In fact, we can write in general $\eta = 1/\sqrt{j}$, being $j$ the correlation indexes, in the present case an even number, (in our case $j$=2 and 4).

Hence, from equation (22), we have the following energy states for every electron

$$\frac{E}{\hbar\omega_0} = \frac{1}{2}\left(2n + 1 \pm \frac{1}{j}\right) \tag{23}$$

In Table II we detail these energy states for Landau levels $n$=0, 1 and 2, and values $j$=2 and 4. The fractions already observed experimentally are highlighted in red color. As it is seen, all those values coincide with the families of even denominator, and their corresponding sequences, obtained in the experiments of the FQHE[10, 11, 12, 13, 14, 15].

(The contribution term of the Zeeman and spin-orbit coupling terms to the energy of the electrons of the 2DEG in the case of even denominator states is completely similar to the contribution to the odd denominator states. See above)

All above describes the conditions for the formation of even denominator states and, then, the corresponding plateaux. However, as it shows in the experiments, where it is almost never observed well-defined even denominator states and plateaux, the formation of such kind of states is not easy, what means, in light of the model, that is difficult the formation of such short-range order of cyclotron orbits.

**States associated with no correlations among cyclotron orbits**

From equation (14) is also possible to explain the origin of the fundamental even denominator states and plateaux: {1/2, 3/2, 5/2, 7/2}.

Additionally to the states described in Tables I and II, there are other possible states associated with the lack of correlation among the cyclotron orbits of the electrons, and that we have not considered yet. However. in our framework such states are also included taking into account a correlation index $p$=0 in equation (14), Table III.

A special analysis arises for the case of the 1/2 state, *a priori* expecting to be one of the most important plateaux, but being missing in all the experiments. This has been one of the most intriguing questions related to the FQHE. But, in the scenario describes is possible to understand the reason for that experimental fact: From Table III the 1/2 state is associated with the lack of correlation in the Landau level $n$=0 of the 2DEG, but this state should always appear at values of the magnetic field for what the Larmor radius are so small that the short range order due to the ionized impurities necessarily affecting them, and therefore the 1/2 state, associated with no correlation between each electron and the ionized impurities in the 2DEG, cannot form.

**Summary and Conclusions**

We have presented a different approach to the FQHE, being the change in the symmetry of the Hamiltonian of an electron of any 2DEG, when the magnetic field is increasing in the presence of the electrostatic potential due to the ionized impurities, the responsible of it. As the magnetic field increases the effect of that electrostatic potential evolves; changing in turn the spatial symmetry of the Hamiltonian: from continuous to discrete one. And thus, for example, it is shown that the main series of fractions observed in the experiments, ({1/2, 2/5, 4/9, 5/11, 6/13}, {2/3, 3/5, 4/7, 5/9, 6/11, 7/13}), are a consequence of the breaking of the degeneration of the first Landau level, $n$=0, due to that change in the Hamiltonian symmetry. On one hand, as is hoped for, looking at Figure 1, we see that the most odd probable states (and, then, plateaux) observed correspond to the correlation indexes $p$=1 and $p$=3 (see above for details).

On the other hand, the even states, and the corresponding plateaux, are consequence of the correlations in the electron cyclotron orbits when correlations $j$=2 and 4 (see above for details).

Therefore we think the model justifies all the odd and even plateaux observed in the experiments. We hope that the presented approach could be a good starting point to understand and analyze other quantum Hall effects.

## References

1. Klitzing, K. v., Dorda, G. & Pepper, M. New method for high-accuracy determination of the fine structure constant based on quantized Hall resistance. *Phys. Rev. Lett.* **45**, 494-497 (1980).

2. Laughlin, R. B. Quantized Hall conductivity in two-dimensions. *Physical Review* B **23,** 5632-5633 (1981).

3. Tsui, D. C., Störmer, H. L., Gossard, A. C. Two-dimensional magnetotransport in the extreme quantum limit. *Phys. Rev. Lett.* **48**, 1559-1562 (1982).

4. Prange, R. E., Girvin, S. M. Editors. The Quantum Hall effect. Springer-Verlag (1990).

5. Laughlin, R. B. Anomalous quantum Hall effect: An incompressible quantum fluid with fractionally charged excitations. *Phys. Rev. Lett.* **50**, 1395-1398 (1983).

6. Jain, J. K. Microscopic theory of fractional quantum Hall effect. Advances in Physics **41**, 105-146 (1992).

7. Hidalgo, M. A. Contribución al estudio de los efectos galvanomagnéticos en el gas de electrones bidimensional. *PhD Thesis*. Editorial de la Universidad Complutense de Madrid (1995).

8. Hidalgo, M. A. A semiclassical approach to the integer quantum Hall problem. *Microelectronic Engineering* **43-44**, 453-458 (1998).

9. Hidalgo, M. A., Cangas, R. A model for the study of the Shubnikov-de Haas and the integer quantum Hall effects ina a two dimensional electronic system. *Arxiv*: 0707.4371 (2007)

10. Clark, R. G., Nicholas, R. J., Usher, A., Foxon, C. T., Harris, J. J. Odd and even fractionally quantized states in GaAs-GaAlAs hetrojunctions. *Surface Science* **170**, 141-147 (1986).

11. Willet, R. L. *et al*. Observations of an even-denominator quantum number in the fractional quantum Hall effect. *Phys. Rev. Lett.* **59**, 1776-1779 (1987).

12. Willet, R. L. *et al*. Termination of the series of fractional quantum Hall states at small filling factors. *Phys. Rev. B* **59**, 7881-7884 (1988).

13. Du, R. R., Störmer, H. L., Tsui, D. C., Pfeiffer, L. N., West, K. W. Experimental evidence for new particles in the fractional quantum Hall effect. *Phys. Rev. B* **70**, 2944-2947 (1988).

14. Choi, H. C., Kang, W., Das Sarma, S., Pfeiffer, L. M., West, K. W. Fractional quantum Hall effect in the second Landau level. *Arxiv*: 0707.0236v2 (2007).

15. Shabani, J., Shayegan, M. Fractional quantum Hall effect at high fillings in a two-subband electron system. *Arxiv*: 1004.09/9v1 (2010).

16. Ando, T., Fowler, A. B., Stern, F. Electronic Properties of Two-dimensional Systems *Reviews of Modern Physics* **54**, 437-672 (1982).

**Acknowledgments:**

The author would like to thank R. Cangas for valuable discussions.

**Table I: Fractional energy states for correlation indexes *p*=1 and *p*=3.**

| $\frac{E}{\hbar\omega_0}$ | $\frac{1}{2}\left(2n+1-\frac{1}{q}\right)$ | | | $\frac{1}{2}\left(2n+1+\frac{1}{q}\right)$ | | | $\frac{1}{2}\left(2n+1-\frac{3}{q}\right)$ | | | $\frac{1}{2}\left(2n+1+\frac{3}{q}\right)$ | | |
|---|---|---|---|---|---|---|---|---|---|---|---|---|
| | $n$=0 | $n$=1 | $n$=2 | $n$=0 | $n$=1 | $n$=2 | $n$=0 | $n$=1 | $n$=2 | $n$=0 | $n$=1 | $n$=2 |
| $q$=3 | $\frac{1}{3}$ | $\frac{4}{3}$ | $\frac{7}{3}$ | $\frac{2}{3}$ | $\frac{5}{3}$ | $\frac{8}{3}$ | - | - | - | - | - | - |
| $q$=5 | $\frac{2}{5}$ | $\frac{7}{5}$ | $\frac{12}{5}$ | $\frac{3}{5}$ | $\frac{8}{5}$ | $\frac{13}{5}$ | $\frac{1}{5}$ | $\frac{6}{5}$ | $\frac{11}{5}$ | $\frac{4}{5}$ | $\frac{9}{5}$ | $\frac{14}{5}$ |
| $q$=7 | $\frac{3}{7}$ | $\frac{10}{7}$ | $\frac{17}{7}$ | $\frac{4}{7}$ | $\frac{11}{7}$ | $\frac{18}{7}$ | $\frac{2}{7}$ | $\frac{9}{7}$ | $\frac{16}{7}$ | $\frac{5}{7}$ | $\frac{12}{7}$ | $\frac{19}{7}$ |
| $q$=9 | $\frac{4}{9}$ | $\frac{13}{9}$ | $\frac{22}{9}$ | $\frac{5}{9}$ | $\frac{14}{9}$ | $\frac{23}{9}$ | - | $\frac{12}{9}$ | $\frac{21}{9}$ | - | $\frac{15}{9}$ | $\frac{24}{9}$ |
| $q$=11 | $\frac{5}{11}$ | $\frac{16}{11}$ | $\frac{27}{11}$ | $\frac{6}{11}$ | $\frac{17}{11}$ | $\frac{28}{11}$ | $\frac{4}{11}$ | $\frac{15}{11}$ | $\frac{29}{11}$ | $\frac{7}{11}$ | $\frac{18}{11}$ | $\frac{26}{11}$ |
| $q$=13 | $\frac{6}{13}$ | $\frac{19}{13}$ | $\frac{33}{13}$ | $\frac{7}{13}$ | $\frac{20}{13}$ | $\frac{32}{13}$ | $\frac{5}{13}$ | $\frac{18}{13}$ | $\frac{34}{13}$ | $\frac{8}{13}$ | $\frac{21}{13}$ | $\frac{31}{13}$ |

The fractions already observed in the experiments of literature are highlighted in red[10, 11, 12, 13, 14, 15]. $n$ corresponds to the Landau levels. As it is seen, the main odd denominator fractions usually observed are related to correlations indexes $p$=1 and $p$=3, i.e. correlations lengths with $\eta = 1$ and $\eta = \sqrt{3}$, respectively. It is hoped for observing the fractions detailed in black in future experiments.

**Table II: Fractional energy states for correlation indexes *j*=2 and 4**

The fractions already observed in the experiments of literature are highlighted in red[10, 11, 12, 13, 14, 15]. *n* corresponds to the Landau levels. As it is seen, the main even denominator fractions usually observed are related to correlations indexes *j*=2 and 4, i.e. correlations lengths with $\eta = 1/\sqrt{2}$ and $\eta = 1/2$ It is hoped for observing in future experiments the fractions detailed in black.

| $\frac{E}{\hbar\omega_0}$ | $\frac{1}{2}\left(2n+1-\frac{1}{j}\right)$ | | | | $\frac{1}{2}\left(2n+1+\frac{1}{j}\right)$ | | | |
|---|---|---|---|---|---|---|---|---|
| | *n*=0 | *n*=1 | *n*=2 | *n*=3 | *n*=0 | *n*=1 | *n*=2 | *n*=3 |
| *j*=2 | $\frac{1}{4}$ | $\frac{5}{4}$ | $\frac{9}{4}$ | $\frac{13}{4}$ | $\frac{3}{4}$ | $\frac{7}{4}$ | $\frac{11}{4}$ | $\frac{15}{4}$ |
| *j*=4 | $\frac{3}{8}$ | $\frac{11}{8}$ | $\frac{19}{8}$ | $\frac{27}{8}$ | $\frac{5}{8}$ | $\frac{13}{8}$ | $\frac{21}{8}$ | $\frac{29}{8}$ |

**Table III: Fractional energy states for correlation index *p*=0**

The fractions already observed in the experiments of literature are highlighted in red[10, 11, 12, 13, 14, 15]. *n* corresponds to the Landau levels. As it is seen, the main even denominator fractions observed, or expected to be observed, are due to states corresponding with the lack of correlations among electrons, what in our model means a correlation index *p*=0.

| $\frac{E}{\hbar\omega_0}$ | $\frac{1}{2}(2n+1)$ | | | |
|---|---|---|---|---|
| | *n*=0 | *n*=1 | *n*=2 | *n*=3 |
| - | $\frac{1}{2}$ | $\frac{3}{2}$ | $\frac{5}{2}$ | $\frac{7}{2}$ |

**Figure legends:**

**Figure 1: Short-range order expected in an ionized impurities distribution, and the corresponding arrangement of cyclotron orbits**

Expected close packing arrange of identical electron cyclotron orbits of a 2DEG in fractional quantum Hall conditions. In correspondence with Table I, the correlations detailed in the picture correspond to the correlation indexes $p$=1 and $p$=3, i.e. correlation lengths $\eta = 1$ and $\eta = \sqrt{3}$, respectively. With black points the most probable arrange of ionizes impurities forming a close-packing structure are also highlighted.

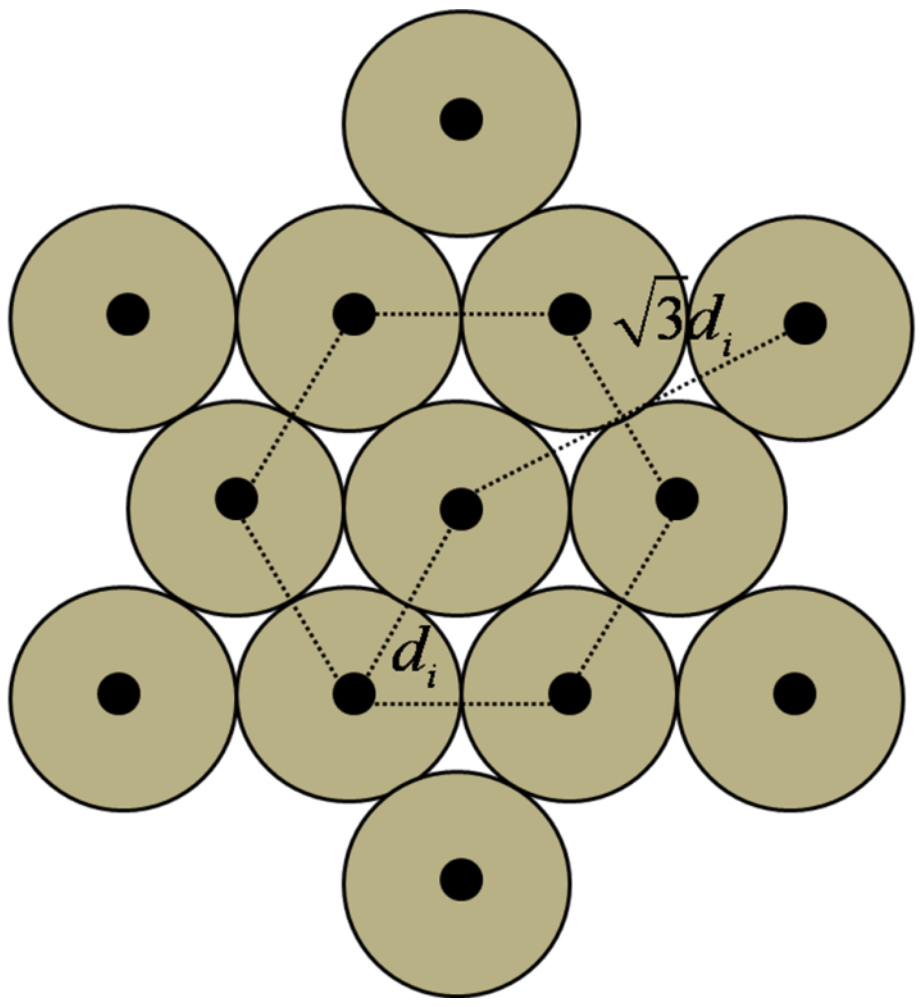

**Figure 2: Simulation with the model just described of the Hall magnetoconductivity and both magnetoresistivities for the set of the odd denominator states *q*=3 and *q*=5, and the corresponding plateaux, both denominators related to a 2DEG sample with $n$=3×10$^{-15}$ m$^{-2}$.**

Taking the experiment by Willet et al.[11] as the reference we show the simulations for the set of states corresponding to the odd denominators a) *q*=3; b) *q*=5. As can be checked, in both cases, the maxima of the Shubnikov-de Haas oscillations and the plateaux of the Hall effect of the model appear at the expected magnetic field values observed in the experimental measurements referenced.

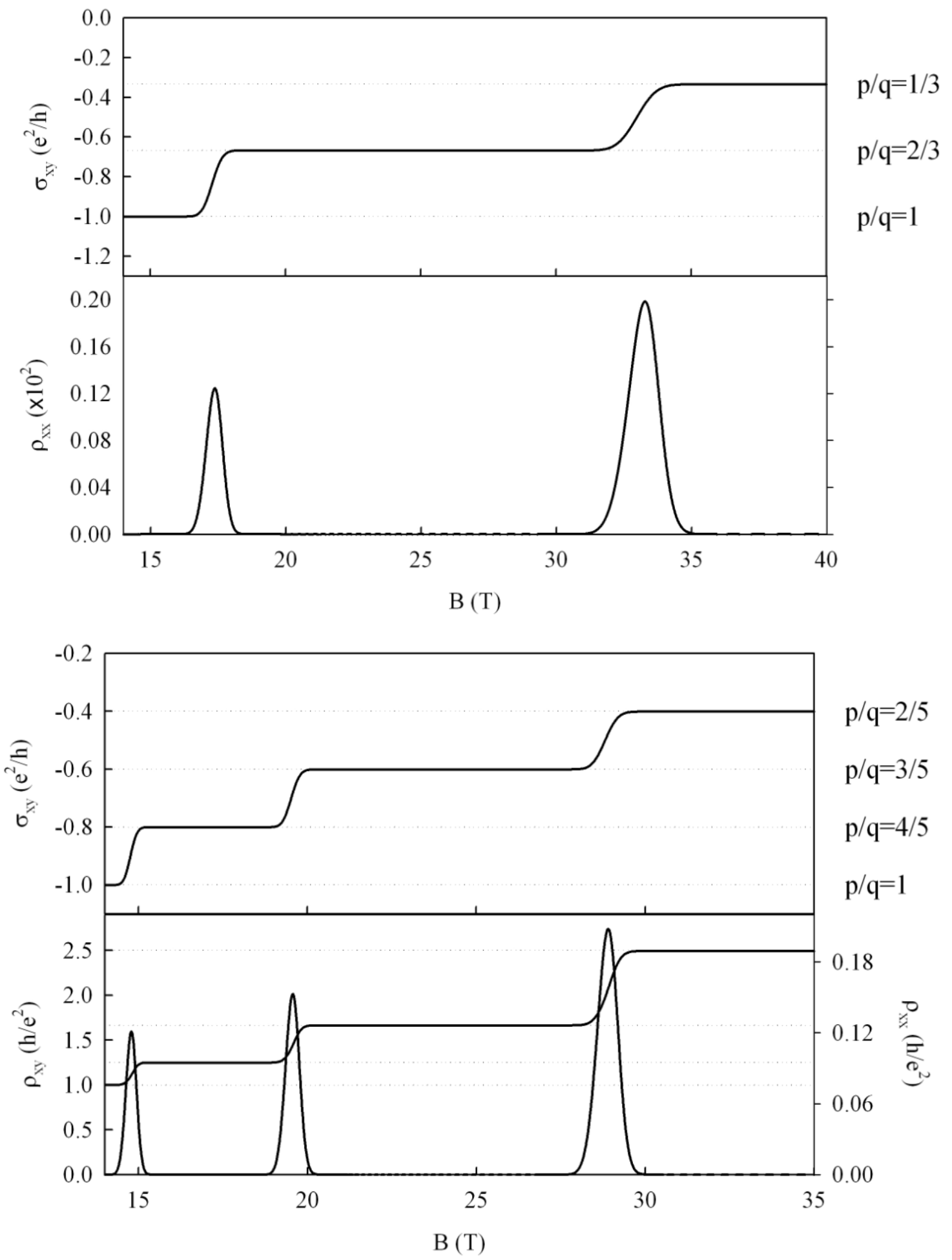

**Figure 3: Short-range order arrangement of the identical electron cyclotron orbits of a 2DEG in even denominator fractional quantum Hall conditions**

Arrangement of identical electron cyclotron orbits of a 2DEG in fractional quantum Hall conditions when the Larmor radius corresponds to the magnetic length. In correspondence with Table II, the correlations detailed in the picture correspond to the correlation indexes usually observed, *j*=2 and 4, i.e. correlations lengths determined by $\eta = 1/\sqrt{2}$ and $\eta = 1/2$. With black points the most probable arrange of ionizes impurities forming a close-packing structure are also highlighted. In blue the cyclotron orbits are represented. The drawn square is a guide for the eyes.

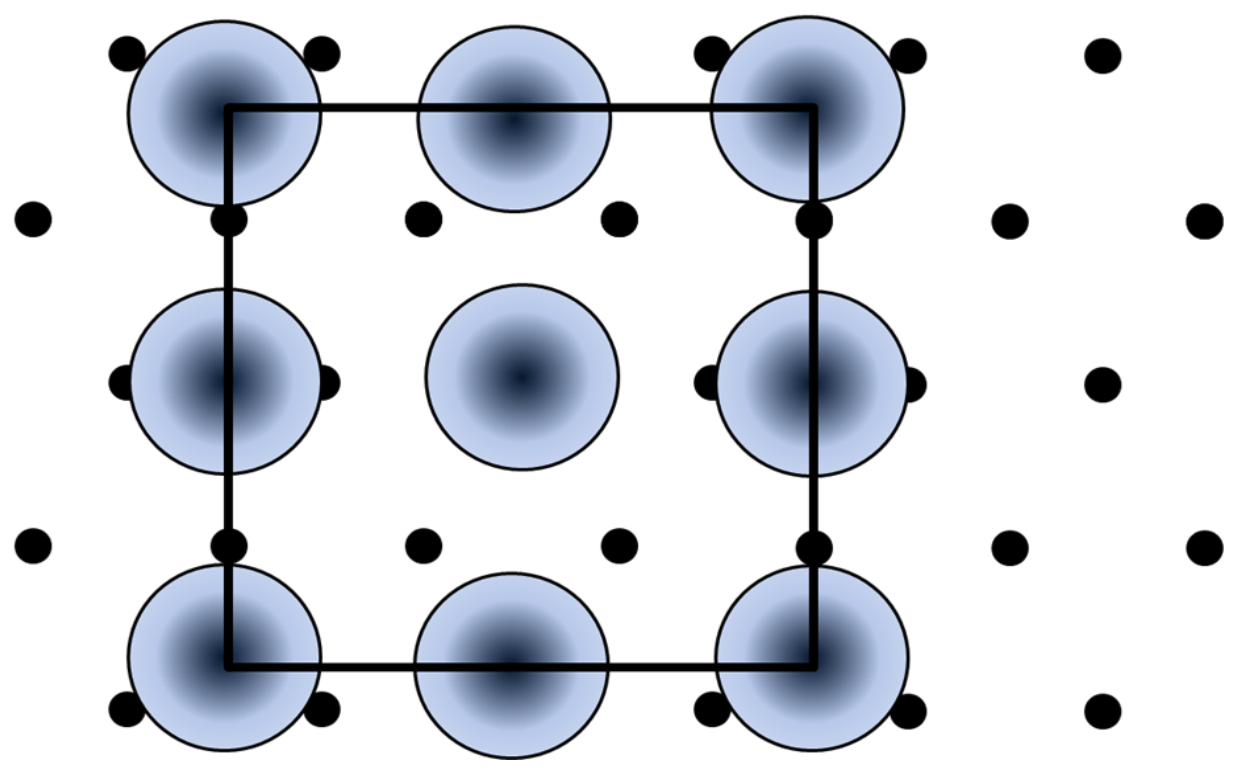